\documentclass[showpacs,preprintnumbers,amsmath,amssymb, prl, twocolumn]{revtex4-1}

\usepackage{stmaryrd}
\usepackage{txfonts}
\usepackage{amssymb}
\usepackage{mathrsfs}
\usepackage{graphicx}
\usepackage{dcolumn}
\usepackage{bm}
\usepackage{epsfig}
\begin{document}

\preprint{S. -Z. Lin and X. Hu, New J. Phys. {\bf{14}}, 063021 (2012)}

\title{Phase solitons in multi-band superconductors with and without time-reversal symmetry}

\author{Shi-Zeng Lin\(^{1,2}\) and Xiao Hu\(^{1}\)}

\affiliation{\(^{1}\) International Center for Materials Nanoarchitectonics (WPI-MANA), National Institute for Materials Science, Tsukuba 305-0044, Japan\\
\(^{2}\)Theoretical Division, Los Alamos National Laboratory, Los Alamos, New Mexico 87545, USA}

\begin{abstract}
The Josephson-like interband couplings in multi-band
superconductivity exhibit degenerate energy minima, which support
states with kinks in phase of superconductivity. When the
interband couplings in systems of three or more components are
frustrated, the time-reversal symmetry (TRS) can be broken, which
generates another type of phase kink between the two
time-reversal-symmetry breaking (TRSB) pair states. In this work, we
focus on these novel states of phase kinks, and investigate their
stability, similarity, differences and physical consequences. Main
results are summarized as follows: (1) We find a new type of phase
slip when the kink becomes unstable. (2) In the kink region, TRS is
broken and spontaneous magnetic fields are induced. (3) In
superconductors with TRSB, composite topological excitations
associated with variations of both superconductivity phase and
amplitude can be created by local perturbations, or due to proximity
effect between normal metals.
\end{abstract}


\date{\today}

\maketitle \noindent {\it Introduction --} It has been known for
long time that superconductors with different pairing symmetries in
contact with one another can form stable domain
structures\cite{Tsuei94,Ng09}. Properties of domain walls are
governed by the pairing symmetries in the domains, thus these
heterogeneous systems become vital in understanding the pair
symmetry. Meanwhile, there are growing evidences that
superconductors may break discrete symmetries in addition to the
$U(1)$ (local) gauge symmetry whose loss defines
superconductivity\cite{Sigrist91rmb,Sigrist91}. Examples include
TRSB in some unconventional superconductors\cite{Volovik85,Lee09},
which results in unusual phenomena such as the appearance of
magnetic flux when the superconductivity is perturbed by nonmagnetic
impurities\cite{Choi89}. These superconductors can also form stable
domain walls between domains of distinct symmetry-breaking states.

The discovery of $\rm{MgB_2}$\cite{Nagamatsu01} and iron-pnictide
superconductors\cite{Kamihara08} has opened intense and
exciting discussions of multi-band superconductivity in condensed
matter physics. In these systems, superconductivity in one band is
coupled through interband Josephson coupling to that in another band
$\gamma_{ij}\Delta_i\Delta_j\cos (\phi_i-\phi_j)$, with $\phi_i$ and
$\Delta_i$ being the superconductivity phase and amplitude in the
$i$-th band respectively. This gives rise a collective oscillation
of the superconductivity phases, known as the Leggett
mode\cite{Leggett66}.

It is interesting to observe that the interband coupling has
degenerate energy minima $\phi_i-\phi_j=2n\pi$ for $\gamma_{ij}<0$,
which supports various topological excitations in the form of phase kinks
belonging to the homotopy class $\pi_0(S^0)$, whereas the well known
vortex solution in type II superconductors belongs to the homotopy
class $\pi_1(S^1)$. The existence of the kink solution was first
discussed by Tanaka\cite{tanaka02} for two-component
superconductors, and later it was discussed that phase kinks can be
excited in nonequilibrium processes such as current
injection\cite{Gurevich03}. The phase kinks have been observed experimentally in layered aluminium mesocopic rings with two order parameters\cite{Bluhm06}. 

In the presence of frustrated interband couplings in superconductors
with three or more components, the system may break the
TRS\cite{Agterberg99,Stanev10,Tanaka10,Hu11, Lin11Leggett}. In the
TRSB state, $\hat\Psi\neq e^{i\theta}\hat\Psi^*$ for any phase
$\theta$ with $\hat\Psi\equiv (\Psi_1, \Psi_2, ..., \Psi_n)$ a
vector of the complex order parameters. A phase kink may appear
between two degenerate states $\hat\Psi$ and $\hat\Psi^*$. One thus
sees that multi-band superconductors support two types of kink
solutions of different origins. In a recent paper by Garaud
\emph{et. al.}\cite{Garaud11}, composite topological excitations
associated with phase kink and vortex in superconductors with TRSB
have been found numerically. The stability of these phase
kinks however still remains to be investigated.

In the present work, we investigate the stability, similarity,
differences and physical consequences of these two types of kink
solutions. In superconductors with TRS in bulk, the phase kink
breaks TRS at the domain wall, and induces local magnetic flux. Upon
elevation of temperatures, kinks become unstable as a consequence of
increasing coherence length. At the instability, a phase slippage
occurs accompanying a voltage pulse. Contrarily, kinks between TRSB
pair states remain stable even with the increasing coherence length.
Moreover, in superconductors with TRSB, various types of composite
topological excitations associated with the variation of
superconductivity phase and amplitude can be created by perturbing
the superconductivity locally, such as heating and/or nonmagnetic
impurities, and at interface to a normal metal.

\vspace{3mm}
\noindent {\it Kink solutions --} We start from the standard multi-band Ginzburg-Landau (GL) theory with Josephson-like interband couplings\cite{Zhitomirsky04,Gurevich07}, which is adequate for discussions on physics addressed here
\begin{equation}\label{eq1}
\begin{array}{l}
{\cal F} = \sum\limits_j {\left[ {{\alpha _j}{{\left| {{\Psi _j}} \right|}^2} + \frac{{{\beta _j}}}{2}{{\left| {{\Psi _j}} \right|}^4} + \frac{1}{{2{m_j}}}{{\left| {\left( { - i \nabla  - {\bf{A}}} \right){\Psi _j}} \right|}^2}} \right]} \\
 + \frac{1}{{8\pi }}{(\nabla  \times {\bf{A}})^2} + \sum\limits_{l \ne j} {{\gamma _{lj}}} \left( {{\Psi _l}\Psi _j^* + {\rm{c}}{\rm{.c}}{\rm{.}}} \right),
\end{array}
\end{equation}
where symbols are conventionally defined\cite{TinkhamBook}.  Throughout the paper, we use the units $\hbar=2e=c=1$. $\gamma_{lj}$ for $l\neq j$ is the interband coupling, which can be either repulsive or attractive depending on the strengths of the Coulomb and electron-phonon interactions. The interband repulsion may cause frustration of the superconductivity in different bands and results in TRSB\cite{Agterberg99,Stanev10,Tanaka10,Hu11, Lin11Leggett}. For $\rm{MgB_2}$, the interband coupling is commonly accepted as attractive $\gamma_{12}<0$, while for iron-pnictide superconductors, there are growing evidences that some of $\gamma_{ij}$ are positive and the system favors $s\pm$ pair symmetry.\cite{Mazin08,Kuroki08}

\begin{figure}[t]
\psfig{figure=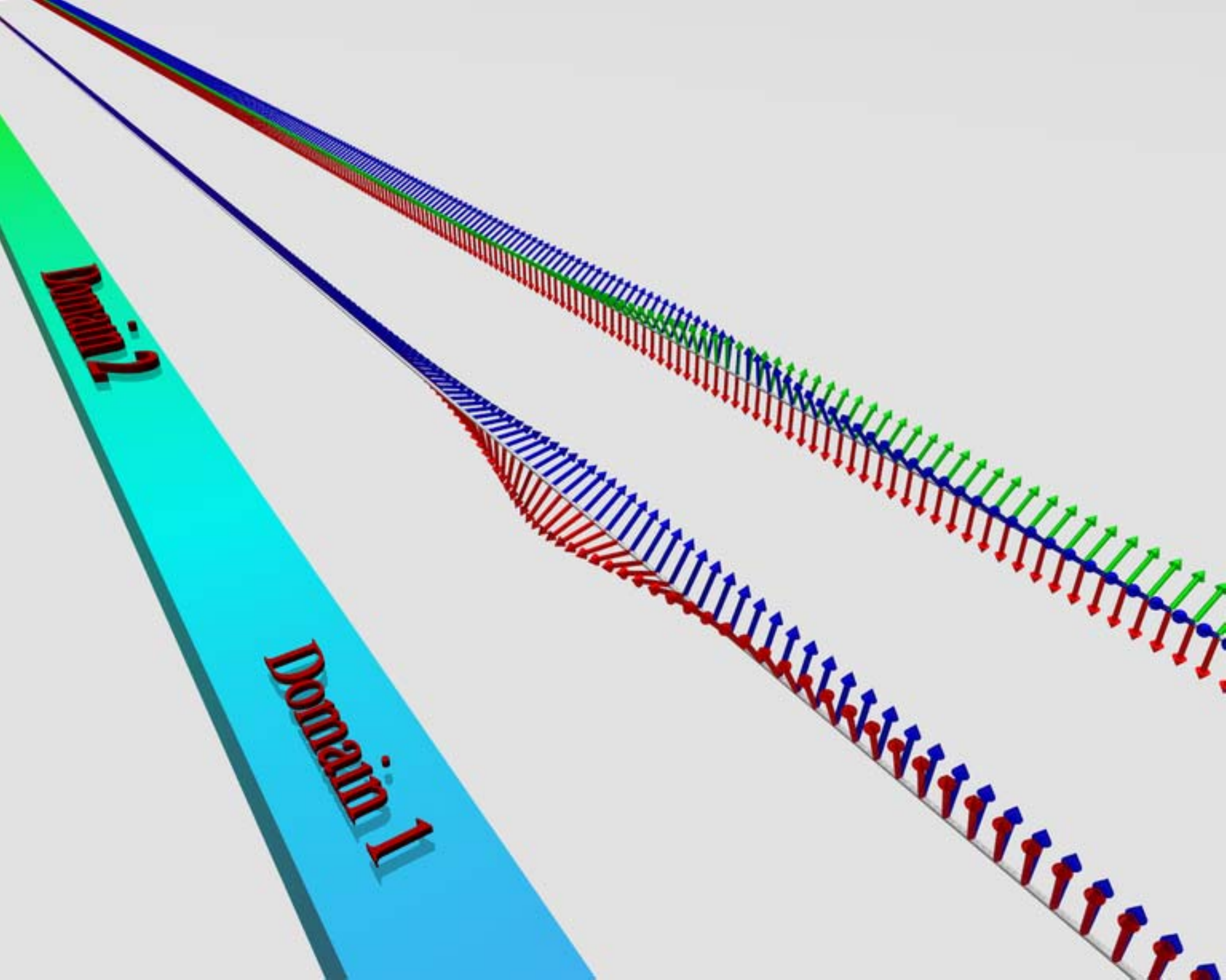,width=\columnwidth} \caption{\label{f0} (color online). Domain structure in multi-band superconductors (left), phase kink in a two-band superconductor (middle), and that in a three-band superconductor with each domain corresponds to distinct TRSB pair states (right).}
\end{figure}

 The kinetic energy in Eq. (\ref{eq1}) can be rewritten as $\frac{H_{cj}^2}{4\pi}\xi_j^2 \left|(\nabla+ i  \mathbf{A} )\Psi_j\right|^2$, where $\xi_j=\sqrt{1/2m_j |a_j|}$ and $H_{cj}$ are coherence length and thermodynamic critical field in respective single-band condensates ($\gamma_{lj}=0$). When width of the kink $\lambda_k$ (derived below) is much larger than $\xi_j$, $\lambda_k\gg\xi_j$, the suppression of the amplitude of the order parameters by the phase kink is weak, and the order parameter is approximately constant in space. In this case, we can concentrate on the phase variables of the order parameters.

First we consider phase kink between domains with TRS in one dimension, where we can take the gauge $\mathbf{A}=0$. The minimal model for this domain structure is of two bands. Since the sign of $\gamma_{12}$ can be gauged away in this case, we consider $\gamma_{12}<0$ without loss of generality. We also assume an identical amplitude of order parameter $\Delta_i=\Delta$ for simplicity. The variation of the phase difference $\phi_{12}\equiv\phi_1-\phi_2$ is described by the sine-Gordon equation\cite{tanaka02}
\begin{equation}\label{eq2}
\partial_x ^2\phi _{12}+2\gamma _{12} (m_1+m_2)\sin \phi_{12}=0,
\end{equation}
and $\partial_x\phi_1=-m_1\partial_x\phi_2/(m_1+m_2)$. The width of
the kink is $\lambda_k=1/\sqrt{-2\gamma _{12} (m_1+m_2)}$, which is
temperature independent. The condition that $\lambda_k\gg\xi_j$ then
becomes $|\gamma_{lj}|\ll\alpha_j$. A typical phase kink is shown in
Fig. \ref{f0} (middle). The TRS is broken at the domain wall while
reserved in the domains. There are finite phase differences between
the right and left domains in both components.

\begin{figure}[t]
\psfig{figure=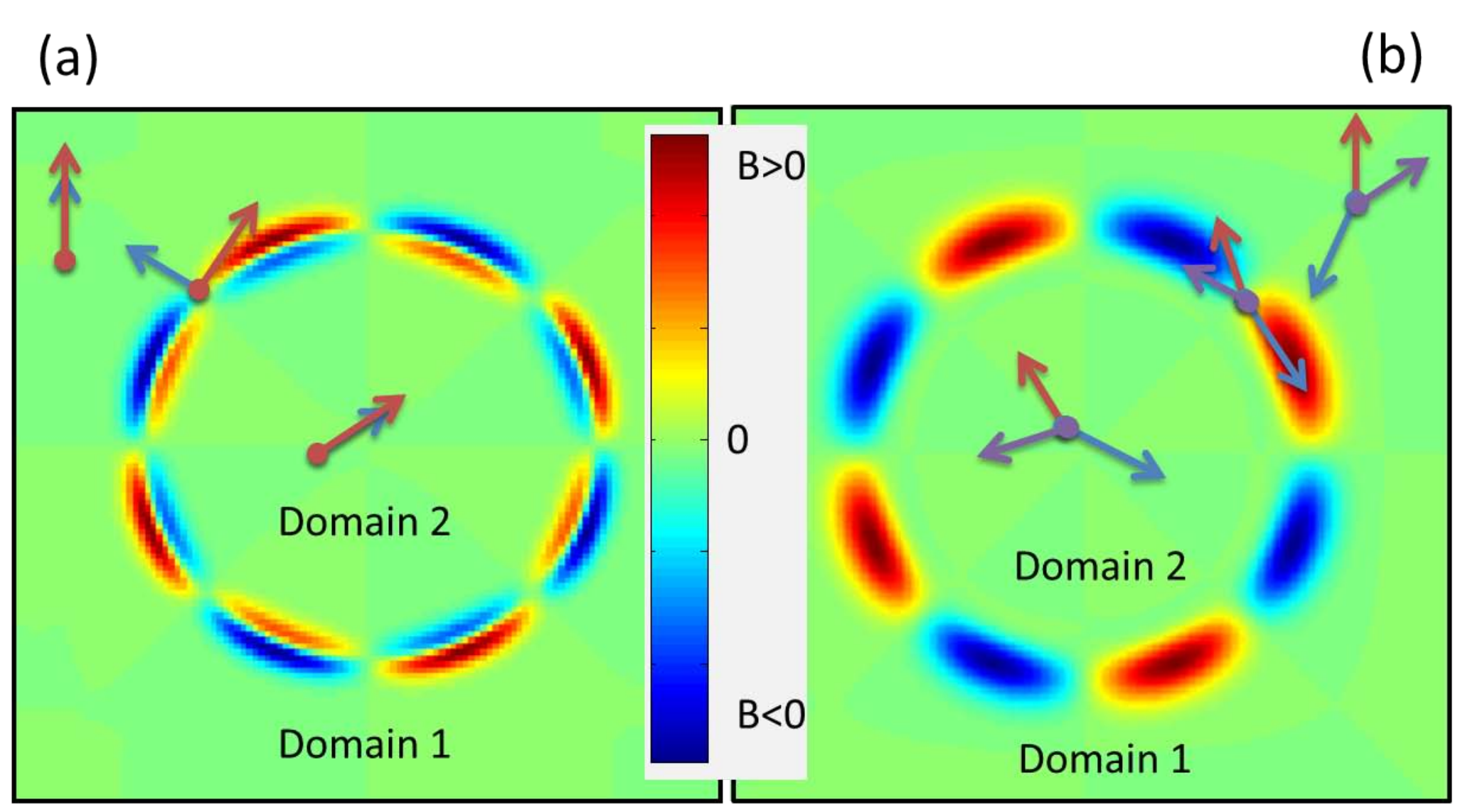,width=\columnwidth} \caption{\label{f1} (color online). Numerical results of the magnetic field distribution: (a) a circular domain wall in a two-band superconductor; (b) a circular domain wall between two TRSB pair states in a three-band superconductor. For (a), $\alpha_j=-20$, $\gamma_{12}=-1$, $\beta_j=1$, $m_1=1$ and $m_2=3$ in the numerical calculations; for (b),  $\alpha_j=0$, $\beta_j=m_j=p_j=1$, $\gamma_{12}=1$, $\gamma_{13}=1.2$ and $\gamma_{23}=1.5$. }
\end{figure}

Now we consider the kink solution of a superconductor of three or more components with frustrated interband couplings, where TRS is broken in bulk. As a minimal model, we treat a superconductor with three equivalent bands $\alpha_{j}=\alpha$, $\gamma_{ij}=\gamma>0$, and $m_i=m$. The two degenerate ground states $\hat\Psi=\Delta (1, e^{i2\pi/3}, e^{i4\pi/3})$ and $\hat\Psi^*=\Delta (1, e^{-i2\pi/3}, e^{-i4\pi/3})$
as a consequence of TRSB are displayed in Fig.\ref{f0} (right). For constant amplitudes of order parameters at $\gamma\ll\alpha$, the phase kink is described by $\partial_x \phi_1=0$, $\partial_x (\phi_{12}+\phi_{13})=0$ and
 \begin{equation}\label{eq3}
\frac{1}{2 m \gamma }\partial_x ^2\phi _{12}+\sin \phi _{12}+\sin \left(2\phi _{12}\right)=0.
\end{equation}
The potential associated with Eq.~(\ref{eq3}) $V_p=\cos\phi_{12}+\cos(2\phi_{12})/2$ has many degenerate minima $\phi_{12,m}=\pm 2\pi/3+2n\pi$. One can construct kink solution between any pair of the energy minima. Their stability and magnetic response are qualitatively the same. To be specific, we only consider the following kink solution, which can be found analytically using the Bogomolny inequality\cite{Manton04}
 \begin{equation}\label{eq4}
\phi _{12}=2 \arctan  \left[\sqrt{3}\text{tanh}\left(-\frac{\sqrt{3 m\gamma }}{2}x\right)\right],
\end{equation}
and the associated energy is
 \begin{equation}\label{eq4a}
E_k=\frac{4}{3}\sqrt{m\gamma} \left(3 \sqrt{3}-\pi \right).
\end{equation}

In one dimension (1D), there is no supercurrent in the domain wall
due to the current conservation $\partial_x J_s=0$. In higher
dimensions, supercurrent and the associated magnetic field are
induced at the domain wall as a result of TRSB. We consider a closed
domain wall described by either Eq. (\ref{eq2}) or Eq.
(\ref{eq4}) in a 2D superconductor. To investigate the dynamic
evolution of the domain wall, we solve the time-dependent GL
equation(TDGL) numerically\cite{szlin10a}
\begin{equation}\label{eq5}
 \frac{{{\hbar ^2}}}{{2{m_j}D_j}}({\partial _t} + i\frac{{{2e}}}{\hbar }\Phi )\Psi_j   =   - \frac{{\delta \mathcal{F}}}{{\delta {\Psi_j ^*}}},
\end{equation}
\begin{equation}\label{eq6}
 \frac{\sigma }{c}(\frac{1}{c}{\partial _t}{\bf{A}} + \nabla \Phi )  =   - \frac{{\delta \mathcal{F}}}{{\delta {\bf{A}}}},
\end{equation}
with $D_j$ the diffusion constant, $\sigma$ the normal conductivity, and $\Phi$ the electric potential.

In simulations, we prepare a closed domain wall with square or rectangular shapes as initial conditions. In order to minimize its energy, the domain wall organizes itself
into a circular shape irrespective to its initial shape during the time evolution in simulations. Magnetic fields appear spontaneously at the
domain wall with alternating directions, as shown in Fig. \ref{f1}
(a) and (b). As revealed by numerical simulations, for
phase kinks in superconductors with TRS [see Eq. (\ref{eq2})], the
induced magnetic field changes polarization in both radial and
azimuthal directions as shown in Fig. \ref{f1}
(a). While for kinks between TRSB pair states [see
Eq. (\ref{eq4})], the magnetic field changes polarization only in
the azimuthal direction as shown in Fig. \ref{f1}
(b). One may treat the domain wall at the left
semicircle as a phase kink, then the domain wall at the right
semicircle is an anti-kink. They attract each other, which causes
the whole circular domain wall collapsing, and renders a uniform
state. Since the attraction between two domain walls becomes
exponentially weak at a large separation, the life time of the
domain walls increases with the size of the domain enclosed. This
allows for possible experimental detections on the induced magnetic
flux after quenching when domain walls are excited by chance.

\vspace{3mm} \noindent {\it Stability of the Kink Solution--} We
proceed to investigate the stability of the kink solution in Eq.
(\ref{eq2}) taking into account the suppression of amplitude of
order parameter by the phase kink. The magnitude of the suppression
depends on the ratio of the kink width $\lambda_k$ to the coherence
length $\xi$ as briefly mentioned above. As the coherence length
increases when temperature is elevated while the width of kink
remains almost unchanged, the superconductivity in the domain wall
will be greatly depleted. At a threshold value, the phase kink loses its stability, and system
evolves into a uniform state. There is a voltage pulse associated with varying magnetic field across the domain wall which is experimentally detectable. This process is a new type
of phase slip, different from that in single-band superconductors
carrying supercurrent close to the critical one, with the latter one
caused by fluctuations.\cite{TinkhamBook}

\begin{figure}[t]
\psfig{figure=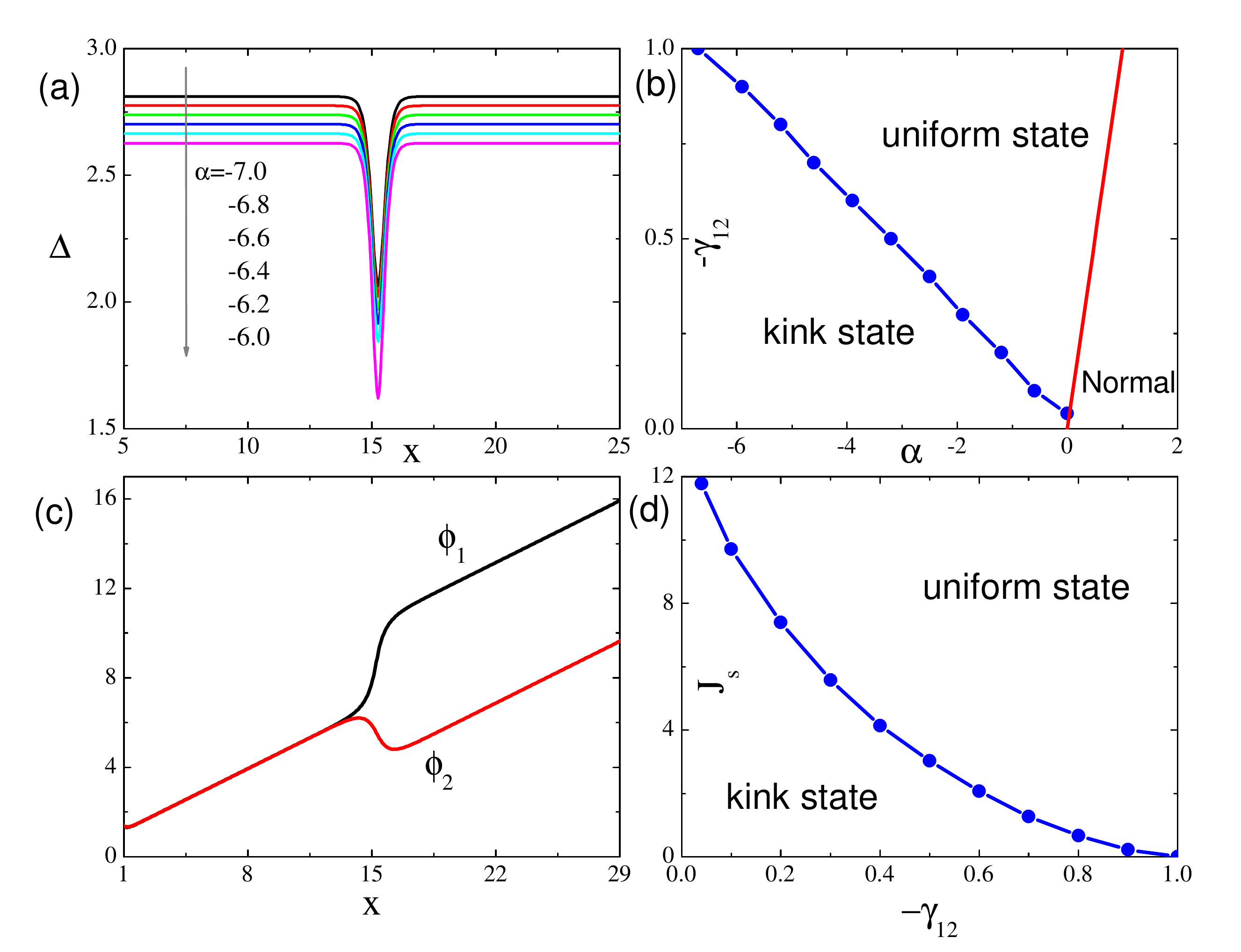,width=\columnwidth} \caption{\label{f2}
(color online). (a) Suppression of the amplitude of
superconductivity at the domain wall when temperature denoted by
$\alpha$ is increased in a two-band superconductor with
$\gamma_{12}=-0.9$. (b) Phase diagram for the stability of phase
kink.  (c) Structure of the phase kink in the presence of
supercurrent. Here $\alpha_j=-7$, $\gamma_{12}=-0.5$ and the
supercurrent $J_s=2.94$. (d) Stability of the phase kink upon
current injection, with $\alpha_j=-7$, $\beta_j=1$ and $m_j=2$.}
\end{figure}

We explicitly consider a superconductor of two identical
bands with a phase kink localized at the center of a superconducting
wire. We solve numerically the TDGL equations and derive the stable
configuration of the superconductivity phase as temperature (namely
$\alpha$) varies. When temperature increases, the amplitude of the
superconductivity at the domain wall decreases as depicted in Fig.
\ref{f2}(a). At a threshold $\alpha$ for given value of
$\gamma_{12}$ [symbols in Fig.~\ref{f2}(b)], the phase kink becomes
unstable and the system evolves into a uniform state, during which a
voltage pulse appears. \emph{Therefore, the phase kinks in superconductors with TRS are stable only
for weak interband couplings.}

A superconducting wire with a phase kink can be alternatively considered as a Josephson junction since the superconductivity is suppressed at the domain wall. In the ground state, the phase difference between two domains is finite, thus it is a realization of $\phi$-junction\cite{Buzdin08}, or $\pi$-junction\cite{Bulaevskii77} if the two bands are identical. When current is injected into the wire, the phase kink is deformed due to the phase gradient created by the injected current. At a threshold current, the phase kink becomes unstable, and the system evolves into a uniform superconducting state, during which a voltage pulse appears, similar to the case with increasing temperature. The threshold current is still much smaller than the depairing current of the uniform state.  One may regard the threshold current as a critical current for the present Josephson junction.

We perform numerical calculations on the critical current of a
superconducting wire with a phase kink, introducing supercurrent
into the system by twisting the phases at the two edges of the wire
far away from the phase kink. The kink structure is deformed into
the shape depicted in Fig. \ref{f2}(c) by the current injection. The
critical current for the kink state decreases with $|\gamma_{12}|$
as shown in Fig.~\ref{f2}(d), since the kink state gradually loses
its stability when $|\gamma_{12}|$ increases as discussed above. At
the critical current, we observe a phase slip with a voltage pulse,
and finally the system reaches a uniform state.

The phase kink in Eq. (\ref{eq4}) between two TRSB pair states of a
three-component superconductor is stable because the system
takes different states in the left and right domains, and one cannot
transform one state to the other by adjusting superconductivity at
the domain wall. This kink is thus protected by symmetry and is very
different from those with bulk TRS as in Eq. (\ref{eq2}), where the
states in the left and right domains are essentially the same except
for a common phase factor, as shown in Fig. \ref{f0} (middle), and
the system evolves into a uniform state by rotating the phase of
domains globally at the instability of the kinks.

\vspace{3mm} \noindent {\it Consequences of TRSB --} We have shown
that spontaneous magnetic flux and voltage pulse appear during the
nonequilibrium evolution of superconductivity phase when a phase
kink becomes unstable. Here we discuss possible experimental
observations on stable phase kinks at \emph{equilibrium}. In the
presence of phase kink, the TRS is violated at the domain wall,
namely $\phi_1-\phi_2\neq 0$ or $\pi$. The variation of the
superconductivity amplitude is coupled with that of
superconductivity phase, which can be checked by expanding the
interband coupling term $\gamma_{ij}\Delta_i\Delta_j\cos
(\phi_i-\phi_j)$ to the quadratic order of phase
difference. When the superconductivity is suppressed locally by
nonmagnetic impurities, proximity effect at sample edge or heating,
variations in superconductivity phases are induced, which in turn
excites supercurrent and magnetic flux, as confirmed numerically.

\begin{figure}[t]
\psfig{figure=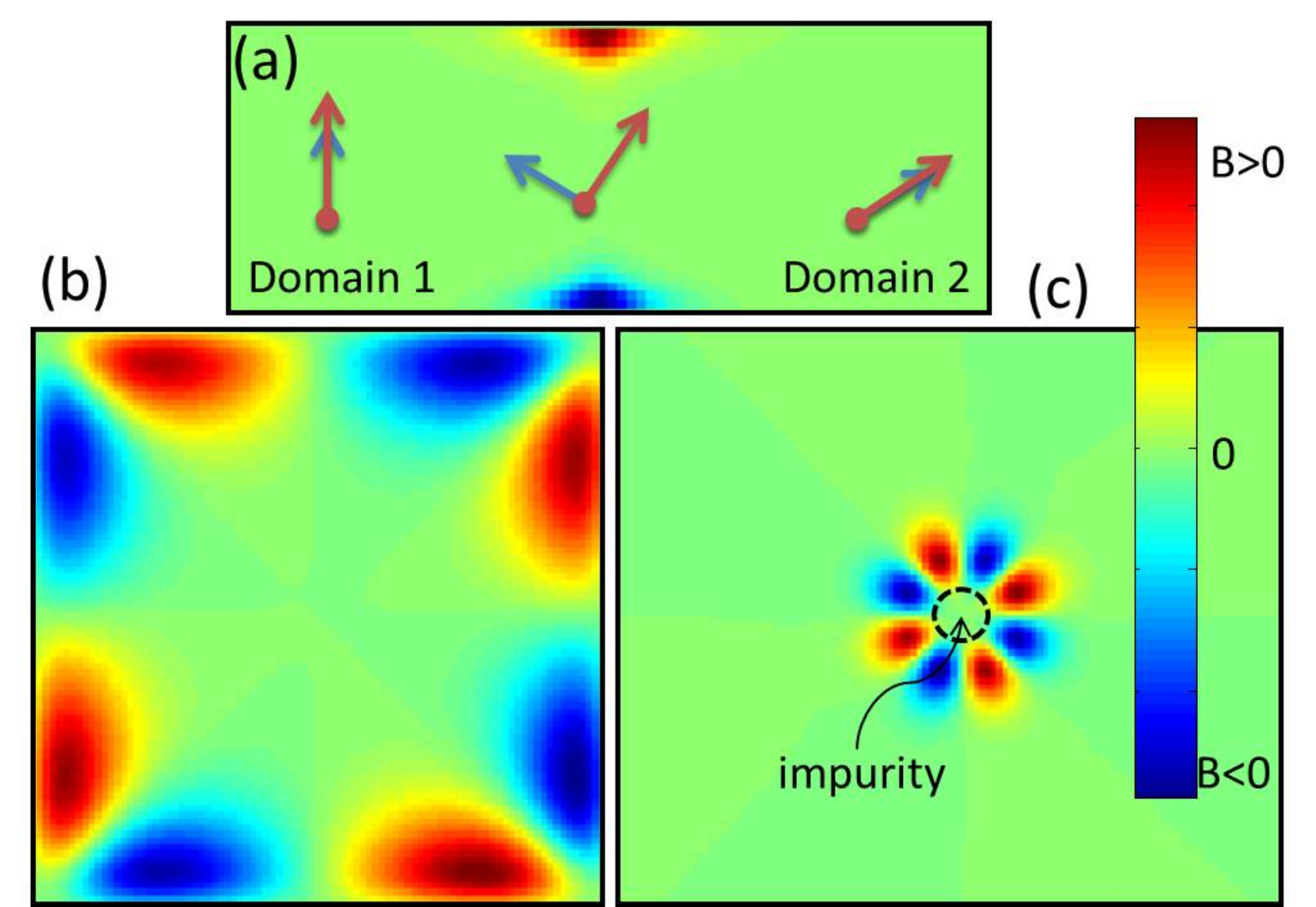,width=\columnwidth} \caption{\label{f3}
(color online). Numerical results of the magnetic field
distribution: (a) a two-band superconducting strip with phase kink
in contact with a normal metal; (b) a three-band superconductor with
TRSB in contact with a normal metal; (c) a three-band superconductor
with TRSB with an impurity. For (a), $\alpha_j=-20$,
$\gamma_{12}=-1$, $\beta_j=1$, $m_1=1$ and $m_2=3$, with proximity
lengths $p_{jj}=2$ and $p_{j\neq l}=\infty$; for (b), $\alpha_j=0$, $\beta_j=m_j=1$, $\gamma_{12}=1$,
$\gamma_{13}=1.2$ and $\gamma_{23}=1.5$, $p_{jj}=1$ and $p_{j\neq
l}=\infty$; for (c), the same as (b) except for $\alpha_j=0.5$
inside the impurity area, and $p_{jl}=\infty$.}
\end{figure}

We study the proximity effect between a superconducting strip and a
normal metal when a phase kink is present in the superconductor. In order to
describe the proximity effect correctly, a boundary condition
between a multi-band superconductor and a normal metals should be
formulated. The boundary condition in terms of the Usadel equation
has been derived in Ref. \cite{Brinkman04}. In the framework of
phenomenological GL theory, the boundary condition for a single-band
superconductor can be generalized straightforwardly to a multi-band
one\cite{TinkhamBook}
\begin{equation}\label{eq7}
\left(-i \nabla -\bf{A}\right)\Psi _j=i\sum_k \frac{\Psi _k}{p_{jk}},
\end{equation}
where the off-diagonal coefficient $p_{jk}$ with $j\neq k$ accounts
for the interband coupling while the diagonal coefficient $j=k$
represents suppression of superconductivity as a consequence of the
leakage of Cooper pairs at the interface. We minimize the GL energy
numerically, and the results are presented in Fig. \ref{f3} (a).
Spontaneous magnetic field is induced at the interface between the
normal metal and superconductor at the position of the domain wall,
which is strong enough (in Fig. \ref{f3}(a), $H\sim 10^{-5}H_{c2}$)
to be measured experimentally by scanning SQUID, Hall, or magnetic
force microscopy. The magnetic field has opposite signs at the two
interfaces, leaving a zero integration over the sample.

In TRSB superconductors, stable domain walls associated with the
variation of superconductivity phase and amplitude can be created by
local perturbations, because the phase is coupled with the amplitude
when TRS is violated. In Fig.~\ref{f3}(b), we consider the proximity
effect between a three-band superconductor with TRSB and a normal
metal. Magnetic fluxes appear at the corners of the superconductor,
associated with sharp changes of phase gradients. In Fig.
~\ref{f3}(c), we introduce an impurity by modifying $\alpha_i$
locally. We see that magnetic flux is induced around the impurity.
For superconductors with TRS, no magnetic filed can be induced by
the proximity effect or impurities, which implies a possible way to
detect the TRSB in experiment.

\vspace{3mm} \noindent {\it Discussions --} In multi-band superconductors with Josephson-like interband coupling, the phase kinks as topological excitations can exist because of the multiple degenerate energy minima associated with the interband Josephson coupling. The phase kink suppresses the superconductivity nearby depending on the ratio of the width of the phase kink to the superconducting coherence length, which causes instability when the suppressed superconductivity is insufficient to maintain the phase coherence for the phase kink. The existence of the phase kink does not require breaking of additional symmetry besides $U(1)$. Instead, the presence of phase kink violates the TRS locally. When bulk multi-band superconductors break TRS, a new type of phase kink can be formed between the two TRSB pair states. The topological solutions (phase kinks) in multi-band superconductors discussed in the present work are different from the topological superconductors realized in materials with strong spin-orbit couplings. 

In 1D, both
phase kink in a superconductor with TRS and that between TRSB pair
states are stable. In 2D, because of the attraction between opposite kinks, the domain
wall collapses and the system reaches a uniform state. This is in
accordance with the Derrick's theorem\cite{Derrick64}, i.e. for an
infinite system, the kink state is only stable in 1D. Kinks can be pinned
by the pinning centers where the superfluid densities are small, since
the loss of superconductivity condensation energy can be reduced by
adapting the domain wall to the pinning centers. This may
prevent the domain wall from collapsing and stabilize the kink in
2D. The kink can also be stabilized when vortices are present as discussed by Garaud \emph{et
al.}\cite{Garaud11}. Domain walls created by local heating or impurities in
superconductors with TRSB are stable in 2D and 3D since they
are enforced by external perturbations.

Let us discuss the realization of the phase kink in Eq. (\ref{eq2}). In iron pnictide superconductors, interband scatterings are
strong\cite{Popovich10}, and thus the phase kinks are unlikely realized.
While for the well-known two-band superconductor $\rm{MgB_2}$ and $\rm{V_3Si}$, it is revealed that interband scatterings
are weak\cite{Golubov02,Kogan09}, which may allow for the excitation of
stable phase kinks in low-temperature region. While for the realization of kink in Eq. (\ref{eq4}), we need multi-band superconductors with TRSB. As discussed in Ref. \cite{Lin11Leggett}, the TRSB state can be achieved by chemical doping in iron pnictide superconductors.  Moreover, phase kinks can also be realized in
hybrid structures with two superconducting films coupled with
Josephson coupling. Recently, Vakaryuk \emph{et al.} proposed to realize the phase kinks in superconductors with the $s\pm$ pairing symmetry by exploiting the proximity effect to a conventional $s$-wave superconductor.\cite{Vakaryuk2012}

\vspace{3mm}
\noindent {\it Acknowledgement --} The authors are
grateful for L. N. Bulaevskii and Z. Wang for discussions. This work
was supported by WPI Initiative on Materials Nanoarchitectonics, and
Grants-in-Aid for Scientific Research (No.22540377), MEXT, Japan,
and partially by CREST, JST. SZL is supported partially by the Los Alamos National
Laboratory under Contract No. E8L5-000100LB.


\begin{thebibliography}{34}%
\makeatletter
\providecommand \@ifxundefined [1]{%
 \@ifx{#1\undefined}
}%
\providecommand \@ifnum [1]{%
 \ifnum #1\expandafter \@firstoftwo
 \else \expandafter \@secondoftwo
 \fi
}%
\providecommand \@ifx [1]{%
 \ifx #1\expandafter \@firstoftwo
 \else \expandafter \@secondoftwo
 \fi
}%
\providecommand \natexlab [1]{#1}%
\providecommand \enquote  [1]{``#1''}%
\providecommand \bibnamefont  [1]{#1}%
\providecommand \bibfnamefont [1]{#1}%
\providecommand \citenamefont [1]{#1}%
\providecommand \href@noop [0]{\@secondoftwo}%
\providecommand \href [0]{\begingroup \@sanitize@url \@href}%
\providecommand \@href[1]{\@@startlink{#1}\@@href}%
\providecommand \@@href[1]{\endgroup#1\@@endlink}%
\providecommand \@sanitize@url [0]{\catcode `\\12\catcode `\$12\catcode
  `\&12\catcode `\#12\catcode `\^12\catcode `\_12\catcode `\%12\relax}%
\providecommand \@@startlink[1]{}%
\providecommand \@@endlink[0]{}%
\providecommand \url  [0]{\begingroup\@sanitize@url \@url }%
\providecommand \@url [1]{\endgroup\@href {#1}{\urlprefix }}%
\providecommand \urlprefix  [0]{URL }%
\providecommand \Eprint [0]{\href }%
\providecommand \doibase [0]{http://dx.doi.org/}%
\providecommand \selectlanguage [0]{\@gobble}%
\providecommand \bibinfo  [0]{\@secondoftwo}%
\providecommand \bibfield  [0]{\@secondoftwo}%
\providecommand \translation [1]{[#1]}%
\providecommand \BibitemOpen [0]{}%
\providecommand \bibitemStop [0]{}%
\providecommand \bibitemNoStop [0]{.\EOS\space}%
\providecommand \EOS [0]{\spacefactor3000\relax}%
\providecommand \BibitemShut  [1]{\csname bibitem#1\endcsname}%
\let\auto@bib@innerbib\@empty
\bibitem [{\citenamefont {Tsuei}\ \emph {et~al.}(1994)\citenamefont {Tsuei},
  \citenamefont {Kirtley}, \citenamefont {Chi}, \citenamefont {Yu-Jahnes},
  \citenamefont {Gupta}, \citenamefont {Shaw}, \citenamefont {Sun},\ and\
  \citenamefont {Ketchen}}]{Tsuei94}%
  \BibitemOpen
  \bibfield  {author} {\bibinfo {author} {\bibfnamefont {C.~C.}\ \bibnamefont
  {Tsuei}}, \bibinfo {author} {\bibfnamefont {J.~R.}\ \bibnamefont {Kirtley}},
  \bibinfo {author} {\bibfnamefont {C.~C.}\ \bibnamefont {Chi}}, \bibinfo
  {author} {\bibfnamefont {L.~S.}\ \bibnamefont {Yu-Jahnes}}, \bibinfo {author}
  {\bibfnamefont {A.}~\bibnamefont {Gupta}}, \bibinfo {author} {\bibfnamefont
  {T.}~\bibnamefont {Shaw}}, \bibinfo {author} {\bibfnamefont {J.~Z.}\
  \bibnamefont {Sun}}, \ and\ \bibinfo {author} {\bibfnamefont {M.~B.}\
  \bibnamefont {Ketchen}},\ }\href@noop {} {\bibfield  {journal} {\bibinfo
  {journal} {Phys. Rev. Lett.}\ }\textbf {\bibinfo {volume} {73}},\ \bibinfo
  {pages} {593} (\bibinfo {year} {1994})}\BibitemShut {NoStop}%
\bibitem [{\citenamefont {Ng}\ and\ \citenamefont {Nagaosa}(2009)}]{Ng09}%
  \BibitemOpen
  \bibfield  {author} {\bibinfo {author} {\bibfnamefont {T.~K.}\ \bibnamefont
  {Ng}}\ and\ \bibinfo {author} {\bibfnamefont {N.}~\bibnamefont {Nagaosa}},\
  }\href@noop {} {\bibfield  {journal} {\bibinfo  {journal} {Europhys. Lett.}\
  }\textbf {\bibinfo {volume} {87}},\ \bibinfo {pages} {17003} (\bibinfo {year}
  {2009})}\BibitemShut {NoStop}%
\bibitem [{\citenamefont {Sigrist}\ and\ \citenamefont
  {Ueda}(1991)}]{Sigrist91rmb}%
  \BibitemOpen
  \bibfield  {author} {\bibinfo {author} {\bibfnamefont {M.}~\bibnamefont
  {Sigrist}}\ and\ \bibinfo {author} {\bibfnamefont {K.}~\bibnamefont {Ueda}},\
  }\href@noop {} {\bibfield  {journal} {\bibinfo  {journal} {Rev. Mod. Phys.}\
  }\textbf {\bibinfo {volume} {63}},\ \bibinfo {pages} {239} (\bibinfo {year}
  {1991})}\BibitemShut {NoStop}%
\bibitem [{\citenamefont {Sigrist}\ \emph {et~al.}(1991)\citenamefont
  {Sigrist}, \citenamefont {Ogawa},\ and\ \citenamefont {Ueda}}]{Sigrist91}%
  \BibitemOpen
  \bibfield  {author} {\bibinfo {author} {\bibfnamefont {M.}~\bibnamefont
  {Sigrist}}, \bibinfo {author} {\bibfnamefont {N.}~\bibnamefont {Ogawa}}, \
  and\ \bibinfo {author} {\bibfnamefont {K.}~\bibnamefont {Ueda}},\ }\href@noop
  {} {\bibfield  {journal} {\bibinfo  {journal} {J. Phys. Soc. Jpn.}\ }\textbf
  {\bibinfo {volume} {60}},\ \bibinfo {pages} {2341} (\bibinfo {year}
  {1991})}\BibitemShut {NoStop}%
\bibitem [{\citenamefont {Volovik}\ and\ \citenamefont
  {Gor'kov}(1985)}]{Volovik85}%
  \BibitemOpen
  \bibfield  {author} {\bibinfo {author} {\bibfnamefont {G.~E.}\ \bibnamefont
  {Volovik}}\ and\ \bibinfo {author} {\bibfnamefont {L.~P.}\ \bibnamefont
  {Gor'kov}},\ }\href@noop {} {\bibfield  {journal} {\bibinfo  {journal} {Zh.
  Eksp. Teor. Fiz.}\ }\textbf {\bibinfo {volume} {88}},\ \bibinfo {pages}
  {1412} (\bibinfo {year} {1985})}\BibitemShut {NoStop}%
\bibitem [{\citenamefont {Lee}\ \emph {et~al.}(2009)\citenamefont {Lee},
  \citenamefont {Zhang},\ and\ \citenamefont {Wu}}]{Lee09}%
  \BibitemOpen
  \bibfield  {author} {\bibinfo {author} {\bibfnamefont {W.~C.}\ \bibnamefont
  {Lee}}, \bibinfo {author} {\bibfnamefont {S.~C.}\ \bibnamefont {Zhang}}, \
  and\ \bibinfo {author} {\bibfnamefont {C.~J.}\ \bibnamefont {Wu}},\
  }\href@noop {} {\bibfield  {journal} {\bibinfo  {journal} {Phys. Rev. Lett.}\
  }\textbf {\bibinfo {volume} {102}},\ \bibinfo {pages} {217002} (\bibinfo
  {year} {2009})}\BibitemShut {NoStop}%
\bibitem [{\citenamefont {Choi}\ and\ \citenamefont {Muzikar}(1989)}]{Choi89}%
  \BibitemOpen
  \bibfield  {author} {\bibinfo {author} {\bibfnamefont {C.~H.}\ \bibnamefont
  {Choi}}\ and\ \bibinfo {author} {\bibfnamefont {P.}~\bibnamefont {Muzikar}},\
  }\href@noop {} {\bibfield  {journal} {\bibinfo  {journal} {Phys. Rev. B}\
  }\textbf {\bibinfo {volume} {39}},\ \bibinfo {pages} {9664} (\bibinfo {year}
  {1989})}\BibitemShut {NoStop}%
\bibitem [{\citenamefont {Nagamatsu}\ \emph {et~al.}(2001)\citenamefont
  {Nagamatsu}, \citenamefont {Nakagawa}, \citenamefont {Muranaka},
  \citenamefont {Zenitani},\ and\ \citenamefont {Akimitsu}}]{Nagamatsu01}%
  \BibitemOpen
  \bibfield  {author} {\bibinfo {author} {\bibfnamefont {J.}~\bibnamefont
  {Nagamatsu}}, \bibinfo {author} {\bibfnamefont {N.}~\bibnamefont {Nakagawa}},
  \bibinfo {author} {\bibfnamefont {T.}~\bibnamefont {Muranaka}}, \bibinfo
  {author} {\bibfnamefont {Y.}~\bibnamefont {Zenitani}}, \ and\ \bibinfo
  {author} {\bibfnamefont {J.}~\bibnamefont {Akimitsu}},\ }\href@noop {}
  {\bibfield  {journal} {\bibinfo  {journal} {Nature}\ }\textbf {\bibinfo
  {volume} {410}},\ \bibinfo {pages} {63} (\bibinfo {year} {2001})}\BibitemShut
  {NoStop}%
\bibitem [{\citenamefont {Kamihara}\ \emph {et~al.}(2008)\citenamefont
  {Kamihara}, \citenamefont {Watanabe}, \citenamefont {Hirano},\ and\
  \citenamefont {Hosono}}]{Kamihara08}%
  \BibitemOpen
  \bibfield  {author} {\bibinfo {author} {\bibfnamefont {Y.}~\bibnamefont
  {Kamihara}}, \bibinfo {author} {\bibfnamefont {T.}~\bibnamefont {Watanabe}},
  \bibinfo {author} {\bibfnamefont {M.}~\bibnamefont {Hirano}}, \ and\ \bibinfo
  {author} {\bibfnamefont {H.}~\bibnamefont {Hosono}},\ }\href@noop {}
  {\bibfield  {journal} {\bibinfo  {journal} {J. Am. Chem. Soc.}\ }\textbf
  {\bibinfo {volume} {130}},\ \bibinfo {pages} {3296} (\bibinfo {year}
  {2008})}\BibitemShut {NoStop}%
\bibitem [{\citenamefont {Leggett}(1966)}]{Leggett66}%
  \BibitemOpen
  \bibfield  {author} {\bibinfo {author} {\bibfnamefont {A.~J.}\ \bibnamefont
  {Leggett}},\ }\href@noop {} {\bibfield  {journal} {\bibinfo  {journal} {Prog.
  Theor. Phys.}\ }\textbf {\bibinfo {volume} {36}},\ \bibinfo {pages} {901}
  (\bibinfo {year} {1966})}\BibitemShut {NoStop}%
\bibitem [{\citenamefont {Tanaka}(2001)}]{tanaka02}%
  \BibitemOpen
  \bibfield  {author} {\bibinfo {author} {\bibfnamefont {Y.}~\bibnamefont
  {Tanaka}},\ }\href@noop {} {\bibfield  {journal} {\bibinfo  {journal} {Phys.
  Rev. Lett.}\ }\textbf {\bibinfo {volume} {88}},\ \bibinfo {pages} {017002}
  (\bibinfo {year} {2001})}\BibitemShut {NoStop}%
\bibitem [{\citenamefont {Gurevich}\ and\ \citenamefont
  {Vinokur}(2003)}]{Gurevich03}%
  \BibitemOpen
  \bibfield  {author} {\bibinfo {author} {\bibfnamefont {A.}~\bibnamefont
  {Gurevich}}\ and\ \bibinfo {author} {\bibfnamefont {V.~M.}\ \bibnamefont
  {Vinokur}},\ }\href@noop {} {\bibfield  {journal} {\bibinfo  {journal} {Phys.
  Rev. Lett.}\ }\textbf {\bibinfo {volume} {90}},\ \bibinfo {pages} {047004}
  (\bibinfo {year} {2003})}\BibitemShut {NoStop}%
\bibitem [{\citenamefont {Bluhm}\ \emph {et~al.}(2006)\citenamefont {Bluhm},
  \citenamefont {Koshnick}, \citenamefont {Huber},\ and\ \citenamefont
  {Moler}}]{Bluhm06}%
  \BibitemOpen
  \bibfield  {author} {\bibinfo {author} {\bibfnamefont {H.}~\bibnamefont
  {Bluhm}}, \bibinfo {author} {\bibfnamefont {N.~C.}\ \bibnamefont {Koshnick}},
  \bibinfo {author} {\bibfnamefont {M.~E.}\ \bibnamefont {Huber}}, \ and\
  \bibinfo {author} {\bibfnamefont {K.~A.}\ \bibnamefont {Moler}},\ }\href@noop
  {} {\bibfield  {journal} {\bibinfo  {journal} {Phys. Rev. Lett.}\ }\textbf
  {\bibinfo {volume} {97}},\ \bibinfo {pages} {237002} (\bibinfo {year}
  {2006})}\BibitemShut {NoStop}%
\bibitem [{\citenamefont {Agterberg}\ \emph {et~al.}(1999)\citenamefont
  {Agterberg}, \citenamefont {Barzykin},\ and\ \citenamefont
  {Gor'kov}}]{Agterberg99}%
  \BibitemOpen
  \bibfield  {author} {\bibinfo {author} {\bibfnamefont {D.~F.}\ \bibnamefont
  {Agterberg}}, \bibinfo {author} {\bibfnamefont {V.}~\bibnamefont {Barzykin}},
  \ and\ \bibinfo {author} {\bibfnamefont {L.~P.}\ \bibnamefont {Gor'kov}},\
  }\href@noop {} {\bibfield  {journal} {\bibinfo  {journal} {Phys. Rev. B}\
  }\textbf {\bibinfo {volume} {60}},\ \bibinfo {pages} {14868} (\bibinfo {year}
  {1999})}\BibitemShut {NoStop}%
\bibitem [{\citenamefont {Stanev}\ and\ \citenamefont
  {Tesanovic}(2010)}]{Stanev10}%
  \BibitemOpen
  \bibfield  {author} {\bibinfo {author} {\bibfnamefont {V.}~\bibnamefont
  {Stanev}}\ and\ \bibinfo {author} {\bibfnamefont {Z.}~\bibnamefont
  {Tesanovic}},\ }\href@noop {} {\bibfield  {journal} {\bibinfo  {journal}
  {Phys. Rev. B}\ }\textbf {\bibinfo {volume} {81}},\ \bibinfo {pages} {134522}
  (\bibinfo {year} {2010})}\BibitemShut {NoStop}%
\bibitem [{\citenamefont {Tanaka}\ and\ \citenamefont
  {Yanagisawa}(2010)}]{Tanaka10}%
  \BibitemOpen
  \bibfield  {author} {\bibinfo {author} {\bibfnamefont {Y.}~\bibnamefont
  {Tanaka}}\ and\ \bibinfo {author} {\bibfnamefont {T.}~\bibnamefont
  {Yanagisawa}},\ }\href@noop {} {\bibfield  {journal} {\bibinfo  {journal} {J.
  Phys. Soc. Jpn.}\ }\textbf {\bibinfo {volume} {79}},\ \bibinfo {pages}
  {114706} (\bibinfo {year} {2010})}\BibitemShut {NoStop}%
\bibitem [{\citenamefont {Hu}\ and\ \citenamefont {Wang}(2012)}]{Hu11}%
  \BibitemOpen
  \bibfield  {author} {\bibinfo {author} {\bibfnamefont {X.}~\bibnamefont
  {Hu}}\ and\ \bibinfo {author} {\bibfnamefont {Z.}~\bibnamefont {Wang}},\
  }\href@noop {} {\bibfield  {journal} {\bibinfo  {journal} {Phys. Rev. B}\
  }\textbf {\bibinfo {volume} {85}},\ \bibinfo {pages} {064516} (\bibinfo
  {year} {2012})}\BibitemShut {NoStop}%
\bibitem [{\citenamefont {Lin}\ and\ \citenamefont {Hu}(2012)}]{Lin11Leggett}%
  \BibitemOpen
  \bibfield  {author} {\bibinfo {author} {\bibfnamefont {S.~Z.}\ \bibnamefont
  {Lin}}\ and\ \bibinfo {author} {\bibfnamefont {X.}~\bibnamefont {Hu}},\
  }\href@noop {} {\bibfield  {journal} {\bibinfo  {journal} {Phys. Rev. Lett.}\
  }\textbf {\bibinfo {volume} {108}},\ \bibinfo {pages} {177005} (\bibinfo
  {year} {2012})}\BibitemShut {NoStop}%
\bibitem [{\citenamefont {Garaud}\ \emph {et~al.}(2011)\citenamefont {Garaud},
  \citenamefont {Carlstr\"{o}m},\ and\ \citenamefont {Babaev}}]{Garaud11}%
  \BibitemOpen
  \bibfield  {author} {\bibinfo {author} {\bibfnamefont {J.}~\bibnamefont
  {Garaud}}, \bibinfo {author} {\bibfnamefont {J.}~\bibnamefont
  {Carlstr\"{o}m}}, \ and\ \bibinfo {author} {\bibfnamefont {E.}~\bibnamefont
  {Babaev}},\ }\href@noop {} {\bibfield  {journal} {\bibinfo  {journal} {Phys.
  Rev. Lett.}\ }\textbf {\bibinfo {volume} {107}},\ \bibinfo {pages} {197001}
  (\bibinfo {year} {2011})}\BibitemShut {NoStop}%
\bibitem [{\citenamefont {Zhitomirsky}\ and\ \citenamefont
  {Dao}(2004)}]{Zhitomirsky04}%
  \BibitemOpen
  \bibfield  {author} {\bibinfo {author} {\bibfnamefont {M.~E.}\ \bibnamefont
  {Zhitomirsky}}\ and\ \bibinfo {author} {\bibfnamefont {V.~H.}\ \bibnamefont
  {Dao}},\ }\href@noop {} {\bibfield  {journal} {\bibinfo  {journal} {Phys.
  Rev. B}\ }\textbf {\bibinfo {volume} {69}},\ \bibinfo {pages} {054508}
  (\bibinfo {year} {2004})}\BibitemShut {NoStop}%
\bibitem [{\citenamefont {Gurevich}(2007)}]{Gurevich07}%
  \BibitemOpen
  \bibfield  {author} {\bibinfo {author} {\bibfnamefont {A.}~\bibnamefont
  {Gurevich}},\ }\href@noop {} {\bibfield  {journal} {\bibinfo  {journal}
  {Physica C}\ }\textbf {\bibinfo {volume} {456}},\ \bibinfo {pages} {160}
  (\bibinfo {year} {2007})}\BibitemShut {NoStop}%
\bibitem [{\citenamefont {Tinkham}(1996)}]{TinkhamBook}%
  \BibitemOpen
  \bibfield  {author} {\bibinfo {author} {\bibfnamefont {M.}~\bibnamefont
  {Tinkham}},\ }\href@noop {} {\emph {\bibinfo {title} {Introduction to
  Superconductivity}}}\ (\bibinfo  {publisher} {McGraw-Hill, Inc.},\ \bibinfo
  {address} {New York},\ \bibinfo {year} {1996})\BibitemShut {NoStop}%
\bibitem [{\citenamefont {Mazin}\ \emph {et~al.}(2008)\citenamefont {Mazin},
  \citenamefont {Singh}, \citenamefont {Johannes},\ and\ \citenamefont
  {Du}}]{Mazin08}%
  \BibitemOpen
  \bibfield  {author} {\bibinfo {author} {\bibfnamefont {I.~I.}\ \bibnamefont
  {Mazin}}, \bibinfo {author} {\bibfnamefont {D.~J.}\ \bibnamefont {Singh}},
  \bibinfo {author} {\bibfnamefont {M.~D.}\ \bibnamefont {Johannes}}, \ and\
  \bibinfo {author} {\bibfnamefont {M.~H.}\ \bibnamefont {Du}},\ }\href@noop {}
  {\bibfield  {journal} {\bibinfo  {journal} {Phys. Rev. Lett.}\ }\textbf
  {\bibinfo {volume} {101}},\ \bibinfo {pages} {057003} (\bibinfo {year}
  {2008})}\BibitemShut {NoStop}%
\bibitem [{\citenamefont {Kuroki}\ \emph {et~al.}(2008)\citenamefont {Kuroki},
  \citenamefont {Onari}, \citenamefont {Arita}, \citenamefont {Usui},
  \citenamefont {Tanaka}, \citenamefont {Kontani},\ and\ \citenamefont
  {Aoki}}]{Kuroki08}%
  \BibitemOpen
  \bibfield  {author} {\bibinfo {author} {\bibfnamefont {K.}~\bibnamefont
  {Kuroki}}, \bibinfo {author} {\bibfnamefont {S.}~\bibnamefont {Onari}},
  \bibinfo {author} {\bibfnamefont {R.}~\bibnamefont {Arita}}, \bibinfo
  {author} {\bibfnamefont {H.}~\bibnamefont {Usui}}, \bibinfo {author}
  {\bibfnamefont {Y.}~\bibnamefont {Tanaka}}, \bibinfo {author} {\bibfnamefont
  {H.}~\bibnamefont {Kontani}}, \ and\ \bibinfo {author} {\bibfnamefont
  {H.}~\bibnamefont {Aoki}},\ }\href@noop {} {\bibfield  {journal} {\bibinfo
  {journal} {Phys. Rev. Lett.}\ }\textbf {\bibinfo {volume} {101}},\ \bibinfo
  {pages} {087004} (\bibinfo {year} {2008})}\BibitemShut {NoStop}%
\bibitem [{\citenamefont {Manton}\ and\ \citenamefont
  {Sutcliffe}(2004)}]{Manton04}%
  \BibitemOpen
  \bibfield  {author} {\bibinfo {author} {\bibfnamefont {N.}~\bibnamefont
  {Manton}}\ and\ \bibinfo {author} {\bibfnamefont {P.~M.}\ \bibnamefont
  {Sutcliffe}},\ }\href@noop {} {\emph {\bibinfo {title} {Topological
  solitons}}}\ (\bibinfo  {publisher} {Cambridge University Press},\ \bibinfo
  {address} {Cambridge},\ \bibinfo {year} {2004})\BibitemShut {NoStop}%
\bibitem [{\citenamefont {Lin}\ and\ \citenamefont {Hu}(2011)}]{szlin10a}%
  \BibitemOpen
  \bibfield  {author} {\bibinfo {author} {\bibfnamefont {S.~Z.}\ \bibnamefont
  {Lin}}\ and\ \bibinfo {author} {\bibfnamefont {X.}~\bibnamefont {Hu}},\
  }\href@noop {} {\bibfield  {journal} {\bibinfo  {journal} {Phys. Rev. B}\
  }\textbf {\bibinfo {volume} {84}},\ \bibinfo {pages} {214505} (\bibinfo
  {year} {2011})}\BibitemShut {NoStop}%
\bibitem [{\citenamefont {Buzdin}(2008)}]{Buzdin08}%
  \BibitemOpen
  \bibfield  {author} {\bibinfo {author} {\bibfnamefont {A.}~\bibnamefont
  {Buzdin}},\ }\href@noop {} {\bibfield  {journal} {\bibinfo  {journal} {Phys.
  Rev. Lett.}\ }\textbf {\bibinfo {volume} {101}},\ \bibinfo {pages} {107005}
  (\bibinfo {year} {2008})}\BibitemShut {NoStop}%
\bibitem [{\citenamefont {Bulaevskii}\ \emph {et~al.}(1977)\citenamefont
  {Bulaevskii}, \citenamefont {Kuzii},\ and\ \citenamefont
  {Sobyanin}}]{Bulaevskii77}%
  \BibitemOpen
  \bibfield  {author} {\bibinfo {author} {\bibfnamefont {L.~N.}\ \bibnamefont
  {Bulaevskii}}, \bibinfo {author} {\bibfnamefont {V.~V.}\ \bibnamefont
  {Kuzii}}, \ and\ \bibinfo {author} {\bibfnamefont {A.~A.}\ \bibnamefont
  {Sobyanin}},\ }\href@noop {} {\bibfield  {journal} {\bibinfo  {journal} {JETP
  Lett.}\ }\textbf {\bibinfo {volume} {25}},\ \bibinfo {pages} {290} (\bibinfo
  {year} {1977})}\BibitemShut {NoStop}%
\bibitem [{\citenamefont {Brinkman}\ \emph {et~al.}(2004)\citenamefont
  {Brinkman}, \citenamefont {Golubov},\ and\ \citenamefont
  {Kupriyanov}}]{Brinkman04}%
  \BibitemOpen
  \bibfield  {author} {\bibinfo {author} {\bibfnamefont {A.}~\bibnamefont
  {Brinkman}}, \bibinfo {author} {\bibfnamefont {A.~A.}\ \bibnamefont
  {Golubov}}, \ and\ \bibinfo {author} {\bibfnamefont {M.~Y.}\ \bibnamefont
  {Kupriyanov}},\ }\href@noop {} {\bibfield  {journal} {\bibinfo  {journal}
  {Phys. Rev. B}\ }\textbf {\bibinfo {volume} {69}},\ \bibinfo {pages} {214407}
  (\bibinfo {year} {2004})}\BibitemShut {NoStop}%
\bibitem [{\citenamefont {Derrick}(1964)}]{Derrick64}%
  \BibitemOpen
  \bibfield  {author} {\bibinfo {author} {\bibfnamefont {G.~H.}\ \bibnamefont
  {Derrick}},\ }\href@noop {} {\bibfield  {journal} {\bibinfo  {journal} {J.
  Mathematical Phys.}\ }\textbf {\bibinfo {volume} {5}},\ \bibinfo {pages}
  {1252} (\bibinfo {year} {1964})}\BibitemShut {NoStop}%
\bibitem [{\citenamefont {Popovich}\ \emph {et~al.}(2010)\citenamefont
  {Popovich}, \citenamefont {Boris}, \citenamefont {Dolgov}, \citenamefont
  {Golubov}, \citenamefont {Sun}, \citenamefont {Lin}, \citenamefont {Kremer},\
  and\ \citenamefont {Keimer}}]{Popovich10}%
  \BibitemOpen
  \bibfield  {author} {\bibinfo {author} {\bibfnamefont {P.}~\bibnamefont
  {Popovich}}, \bibinfo {author} {\bibfnamefont {A.~V.}\ \bibnamefont {Boris}},
  \bibinfo {author} {\bibfnamefont {O.~V.}\ \bibnamefont {Dolgov}}, \bibinfo
  {author} {\bibfnamefont {A.~A.}\ \bibnamefont {Golubov}}, \bibinfo {author}
  {\bibfnamefont {D.~L.}\ \bibnamefont {Sun}}, \bibinfo {author} {\bibfnamefont
  {C.~T.}\ \bibnamefont {Lin}}, \bibinfo {author} {\bibfnamefont {R.~K.}\
  \bibnamefont {Kremer}}, \ and\ \bibinfo {author} {\bibfnamefont
  {B.}~\bibnamefont {Keimer}},\ }\href@noop {} {\bibfield  {journal} {\bibinfo
  {journal} {Phys. Rev. Lett.}\ }\textbf {\bibinfo {volume} {105}},\ \bibinfo
  {pages} {027003} (\bibinfo {year} {2010})}\BibitemShut {NoStop}%
\bibitem [{\citenamefont {Golubov}\ \emph {et~al.}(2002)\citenamefont
  {Golubov}, \citenamefont {Kortus}, \citenamefont {Dolgov}, \citenamefont
  {Jepsen}, \citenamefont {Kong}, \citenamefont {Anderson}, \citenamefont
  {Gibson}, \citenamefont {Ahn},\ and\ \citenamefont {Kremer}}]{Golubov02}%
  \BibitemOpen
  \bibfield  {author} {\bibinfo {author} {\bibfnamefont {A.~A.}\ \bibnamefont
  {Golubov}}, \bibinfo {author} {\bibfnamefont {J.}~\bibnamefont {Kortus}},
  \bibinfo {author} {\bibfnamefont {O.~V.}\ \bibnamefont {Dolgov}}, \bibinfo
  {author} {\bibfnamefont {O.}~\bibnamefont {Jepsen}}, \bibinfo {author}
  {\bibfnamefont {Y.}~\bibnamefont {Kong}}, \bibinfo {author} {\bibfnamefont
  {O.~K.}\ \bibnamefont {Anderson}}, \bibinfo {author} {\bibfnamefont {B.~J.}\
  \bibnamefont {Gibson}}, \bibinfo {author} {\bibfnamefont {K.}~\bibnamefont
  {Ahn}}, \ and\ \bibinfo {author} {\bibfnamefont {R.~K.}\ \bibnamefont
  {Kremer}},\ }\href@noop {} {\bibfield  {journal} {\bibinfo  {journal} {J.
  Phys.: Condens. Matter}\ }\textbf {\bibinfo {volume} {14}},\ \bibinfo {pages}
  {1353} (\bibinfo {year} {2002})}\BibitemShut {NoStop}%
\bibitem [{\citenamefont {Kogan}\ \emph {et~al.}(2009)\citenamefont {Kogan},
  \citenamefont {Martin},\ and\ \citenamefont {Prozorov}}]{Kogan09}%
  \BibitemOpen
  \bibfield  {author} {\bibinfo {author} {\bibfnamefont {V.~G.}\ \bibnamefont
  {Kogan}}, \bibinfo {author} {\bibfnamefont {C.}~\bibnamefont {Martin}}, \
  and\ \bibinfo {author} {\bibfnamefont {R.}~\bibnamefont {Prozorov}},\
  }\href@noop {} {\bibfield  {journal} {\bibinfo  {journal} {Phys. Rev. B}\
  }\textbf {\bibinfo {volume} {80}},\ \bibinfo {pages} {014507} (\bibinfo
  {year} {2009})}\BibitemShut {NoStop}%
\bibitem [{\citenamefont {Vakaryuk}\ \emph {et~al.}(2012)\citenamefont
  {Vakaryuk}, \citenamefont {Stanev}, \citenamefont {Lee},\ and\ \citenamefont
  {Levchenko}}]{Vakaryuk2012}%
  \BibitemOpen
  \bibfield  {author} {\bibinfo {author} {\bibfnamefont {V.}~\bibnamefont
  {Vakaryuk}}, \bibinfo {author} {\bibfnamefont {V.}~\bibnamefont {Stanev}},
  \bibinfo {author} {\bibfnamefont {W.}~\bibnamefont {Lee}}, \ and\ \bibinfo
  {author} {\bibfnamefont {A.}~\bibnamefont {Levchenko}},\ }\href@noop {}
  {\bibfield  {journal} {\bibinfo  {journal} {{arXiv:1203.4554}}\ } (\bibinfo
  {year} {2012})}\BibitemShut {NoStop}%
\end{thebibliography}
%

\end{document}